\documentstyle[11pt,fleqn]{article}
\def\titolo{\par\bigskip\begin{center}\bf\LARGE}
\def\endtitolo{\end{center}\par\bigskip\par\rm\normalsize}
\def\instit{\begin{center}\large}
\def\endinstit{\end{center}\rm\normalsize}
\def\references{\begin{thebibliography}{10}}
\def\endreferences{\end{thebibliography}\end{document}}

\newcommand{\btit}{\begin{titolo}}
\newcommand{\etit}{\end{titolo}}

\renewcommand{\author}[1]{\begin{center}\Large #1\end{center}}
\renewcommand{\date}[1]{\par\bigskip\par\sl\hfill #1\par\medskip\par}

\newcommand{\pacs}[1]{\smallskip\noindent{\sl PACS number(s):
                       \hspace{0.3cm}#1}\par\bigskip}
\newcommand{\babs}{\hrule\par\begin{description}\item{Abstract: }\it}
\newcommand{\eabs}{\par\end{description}\hrule\par\medskip\rm}
\newcommand{\ack}[1]{\par\section*{Acknowledgments} #1}
\newcommand{\ca}[1]{{\cal #1}}         
\newcommand{\hs}{\qquad\qquad}         
\newcommand{\nn}{\nonumber}            
\newcommand{\beq}{\begin{eqnarray}}    
\newcommand{\eeq}{\end{eqnarray}}      
\newcommand{\beqn}{\begin{eqnarray}}   
\newcommand{\eeqn}{\end{eqnarray}}     
\newcommand{\ap}{\left.}               
\newcommand{\at}{\left(}               
\newcommand{\aq}{\left[}               
\newcommand{\ag}{\left\{}              
\newcommand{\cp}{\right.}              
\newcommand{\ct}{\right)}              
\newcommand{\cq}{\right]}              
\newcommand{\cg}{\right\}}             
\newcommand{\R}{\mbox{$I\!\!R$}}                 
\newcommand{\ii}{\infty}                         
\newcommand{\fr}[2]{\mbox{$\frac{#1}{#2}$}}      
\newcommand{\Tr}{\,\mbox{Tr}\,}                  
\newcommand{\al}{\alpha}
\newcommand{\be}{\beta}
\newcommand{\ga}{\gamma}
\newcommand{\de}{\delta}

\newcommand{\ze}{\zeta}

\newcommand{\Ga}{\Gamma}

\newcommand{\La}{\Lambda}

\begin{document}

\begin{center}

{\large \bf THE  EFFECTIVE
ACTION IN GAUGED SUPERGRAVITY ON HYPERBOLIC BACKGROUND AND INDUCED
COSMOLOGICAL CONSTANT}

\vspace{4mm}

{\sc A.A. Bytsenko} \\ {\it Department of Theoretical Physics,
State Technical University, \\ St Petersburg 195251, Russia} \\
{\sc S. D. Odintsov}\footnote{E-mail address: odintsov@ebubecm1.bitnet}
\\
{\it Department E.C.M., Faculty of Physics, University of
Barcelona, \\
Diagonal 647, 08028 Barcelona, Spain} \\
and Pedagogical Institute, 634041, Tomsk, Russia \\
{\sc S. Zerbini} \footnote{E-mail address: zerbini@science.unitn.it}\\
{\it Department of Physics, University of Trento, 38050 Povo,
Italy}    \\ and
{\it I.N.F.N., Gruppo Collegato di Trento}


\vspace{5mm}

\end{center}
\begin{abstract}
The one-loop effective action for 4-dimensional gauged supergravity with
negative cosmological constant, is investigated in space-times
with compact hyperbolic spatial section. The explicit expansion of the
effective action as a power series of the curvature on hyperbolic
background is derived, making use of heat-kernel and
zeta-regularization techniques. The induced  cosmological and Newton
constants are computed.
\end{abstract}

\vspace{8mm}

\pacs{04.60 Quantum theory of gravitation }

One of the basic motivations for studying the quantum gravity on de Sitter
background
\cite{gibb78-138-141,gibp,hawk79b,chri80-170-480,cdrg},
\cite{anti,ford,allen,antm,frad84-234-472,odin89,tayl90-345-210}
has been the fact that it might be a very reasonable candidate for vacuum state
when the classical action contains
a positive cosmological term. It has also been expected that such
investigations  might also provide the framework to solve the cosmological
constant problem
\cite{gibb78-138-141,gibp,chri80-170-480,cdrg,frad84-234-472}.
In particular, in Ref. \cite{tayl90-345-210} a quite interesting model
of Coleman-Weinberg type suppression for the effective cosmological
constant (due to the radiative corrections of Einstein gravity) has
been suggested. One important ingredient of this model has been the
proposal to study the large-distance limit of the quantum gravity
one-loop effective action (for a general review on effective action, see for
example
\cite{buch}) on the de Sitter background. However, the use of such
background has been shown to be inadeguate to implement the
cosmological constant suppression mechanism. With regard to this
issue, the use of an hyperbolic background (i.e. a  gravitational theory with
negative cosmological constant) might slighty improve the situation, as
it has been recently shown  in Ref. \cite{byts94r}, where the pure gravity
has been treated as an example.

There exists a class of theories (gauged supergravities) for which
hyperbolic space-times are the natural vacuum states at, since for
these models  the
effective cosmological constant classically may often be negative.
 It is the purpose of this work to investigate, in the large-distance
limit, the Euclidean one-loop effective action
 for the $O(4)$ gauged supergravity  on a hyperbolic background of the
 form $H^4/\Ga$ (i.e. compact hyperbolic 4-dimensional manifold), making use of
heat-kernel, Selberg trace formula and zeta-function regularization techniques
\cite{hawk77-55-133,dowk76-13-3224,eliz}.
The possibility to realize the
cosmological constant suppression mechanism in the spirit of Ref.
 \cite{tayl90-345-210} is investigated and it is shown that some
impletmentation of it may be achieved.

To start with, let us briefly recall the general formalism enabling
the treat the one-loop effective action. In general, the background geometry
may be
chosen to correspond to constant curvature space:
Euclidean space $\R^4$ with
constant curvature $ k=0 $, the sphere $S^4$ of radius $a$
$(k=a^{-2})$ and the hyperbolic space $H^4$  $(k=-a^{-2})$.
For all of these spaces, the curvature tensor, the Ricci tensor and
the scalar curvature have respectively the form
\beqn
R_{\al \be \ga \de}&=& k\at g_{\al \ga}g_{\be \de}-g_{\al \de}g_{\be
\ga}\ct \,,\nn \\
R_{\al \be}&=& k(N-1)g_{\al \be} \,, \hs R=kN(N-1) \,.
\eeqn
Here we shall mainly deal with the (compact) hyperbolic case.
After a standard functional integration
the Euclidean one-loop effective action may be written
\beq
\Ga^{(1)}=\frac{1}{2}\sum_{p,i} C_p \log \det (O^{(i)}_p/{\mu^2})\,,
\label{2.3}
\eeq
where $\mu^2$ is a normalization parameter, $i=0,1/2,1,3/2,2 $ refer to
scalars,....  transverse symmetric traceless second
rank tensors respectively and $C_p$ are the weights associated with
the Laplace-type operators $O^{(i)}_p $.
However, it is well known that, in the path-integral formulation of
Euclidean quantum gravity, some of these operators are negative.
In the following, we shall assume that, when necessary, the contour
rotations and field redefinitions, in accordance with
the prescriptions of Refs. \cite{gibb78-138-141,polc},
have been done.

 In this paper we shall mainly consider $\La<0$ and
the determinants of the operators $O^{(i)}_p$ will be regularized by means of
the zeta-function technique \cite{hawk77-55-133}. The fact we are
dealing with the case  $\La<0$, implies that all $O^{(i)}_p$ (provided
the integration over imaginary field is performed when necessary) are,
for $a$ sufficiently large, positive definite. Thus, a generic Laplace-type
operator  can be written as
\beq
O_p^{(i)}=a^{-2}L_{p}^{(i)} \,.
\label{10}
\eeq
where  we have introduced the dimensionless operator
$L_{p}^{(i)}  $ of the form
\beq
L_{p}^{(i)} =L^{(i)}+X_p^{(i)}\,, \hs X_p^{(i)}=a^2
|\La_p^{(i)}|+\nu_p^{(i)}\,,
\label{2.4}
\eeq
where $L=-\nabla^2_N$ is the Laplace-Beltrami operator on $M^4$,
$\La_p^{(i)}=b_p^{(i)} \Lambda$, $b_p^{(i)}$ is a non-negative number,
$\nu_p^{(i)}$ are
known constants and we assume $ X_p^{(i)}>0$.
The zeta-function regularization gives
\beq
\log \det \at O_p^{(i)}/{\mu^2} \ct=-\aq \ze'(0|O_p^{(i)})+\log \mu^2
\ze(0|O_p^{(i)})\cq\,,
\label{b}
\eeq
where

\beqn
\zeta(s|O_p^{(i)})&=&\frac{1}{\Ga(s)}\int_0^\ii dt t^{s-1}\Tr
e^{-tO_p^{(i)}}\nn
\\
&=&a^{2s}
\ze(s|L_{p}^{(i)})\,.
\label{2.5}
\eeqn
Now we have
\beq
\ze(z|L_{p}^{(i)})=\frac{1}{\Ga(z)}\int_0^\ii dt t^{z-1}e^{-t X_p^{(i)}}\Tr
e^{-tL}
\label{zeta}
\eeq
On general grounds, one can write the well known heat-kernel expansion,
valid for $t \to 0$
\beq
\Tr e^{-tL} \simeq \sum_rK_rt^{\fr{r-4}{2}}\,,
\label{exp}
\eeq
where the coefficients $K_r$ associated with the operator $L$ can be,
 in principle, computed and depend on the geometry of the compact
manifold. The expansion (\ref{exp}) is valid for a compact smooth manifold with
or without boundary. We shall consider manifolds without boundary,
which a possible presence of conical singularities. For boundaryless manifolds
$K_r=0$  when $r$ is odd.

If we make use of the general heat-kernel expansion, we get
\beq
\zeta(s|O_p^{(i)})=\frac{a^{2s}}
{\Ga(s)}\ag
\sum_r K_{2r}\Ga(s+r-2)X_p^{2-s-r} +J_p(s)^{(i)} \cg\,.
\label{2.8}
\eeq
where $J_p(s)^{(i)}$ are  entire functions of $s$.
The first gamma
functions have a
simple pole at $s=0$. So we have
\beq
\ze(0|O_p^{(i)})=
\sum_{l=0}^2 K_{4-2l}^{(i)}\frac{(-X_p^{(i)})^l}{l!}
\label{21}
\eeq
and
\beqn
\ze'(0|O_p^{(i)})&=&
\sum_{l=0}^2 K_{4-2l}^{(i)}\frac{(-X_p^{(i)})^l}{l!}\at \ga+\Psi(l+1)+\log
(a^2/X_p^{(i)}) \ct \nn\\
&+&  \sum_{r>2}K_{2r}^{(i)}\Ga(r-2)(X_p^{(i)})^{2-r}+J_p^{(i)}(0) \,.
\label{22}
\eeqn

As a result, the one-loop contribution to the effective action  reads
\beqn
\Ga^{(1)}&=&-\frac{1}{2}\sum_{p,i} C_p \ag \log (a^2
\mu^2 /X_p^{(i)}) \ze(0|O_p^{(i)}) +
\sum_{l=0}^2 K_{4-2l}\frac{(-X_p^{(i)})^l}{l!}F(l)  \cp \nn\\
&+& \ap  \sum_{r>2}K_{2r}^{(i)}\Ga(r-2)(X_p^{(i)})^{2-r}+J_{p}^{(i)}(0)  \cg\,,
\label{2.1111}
\eeqn
where $F(0)=0 $ and $ F(l)=\sum_{j=1}^l j^{-1}$.

Up to now, our results have a general form. However, in order to
perform explicit computations the
knowledge of the heat-kernel coefficients $K_{2r}$ and  the
analytical part $J_p(s)$ are necessary. It is well known that the $K_{2r}$
coefficients are, in principle, computable. On the other hand, the
evaluation of the $J_p(s)$ functions requires an analytical continuation and
this is achieved, usually,
throught the explicit knowledge of the spectrum of the operator $L_{p}$.
Unfortunately on  hyperbolic backgrounds (the case under discussion)
the spectrum of the Laplace operator is not explicitly known.
Moreover, the scalar sector may be investigated by making use of Selberg trace
formula techniques. For higher spin fields, the analogue techniques is
more complicated and, to our knowledge, it is not explicitly known.
In such situation, in order to perform explicit calculations, we have
to make use of a further approximation scheme, namely  the
large-distance limit, which is particulary interesting in pure
gravity and it has been proposed in Ref.\cite{tayl90-345-210}
for spherical 4-dimensional background in the discussion of the
cosmological constant issue.

In our language, the large distance limit is equivalent to find the
asymptotics of the effective action for very large $X_p$ (note that
 $b_p$ must be non-vanishing and we left understood that, in the sum over
$p$, the terms corresponding to some $b_p=0$, must be omitted). Thus,
in this limit we can neglect the terms related to $r> 2$ and the
analytical (normally unknown) terms. Thus, retaining only the leading
terms
\beq
\ze(0|O_p^{(i)})&=&\frac{1}{2}
a^4 |\La_p^{(i)}|^2 K_0^{(i)}+a^{2}|\La_p^{(i)}|2 ( \nu_p K_0^{(i)}-K_2^{(i)})
+.... \,,
\label{evenl}
\eeq
\beqn
\ze'(0|O_p^{(i)})&=&\frac{1}{2}
\ag  a^4 |\La_p^{(i)}|^2 ( -\log |\La_p^{(i)}|+
F(2)) K_0^{(i)} \cp \nn  \\
&+& \ap  a^{2}|\La_p^{(i)}| \aq -2\log |\La_p^{(i)}|
  (\nu_p^{(i)} K_0^{(i)}-K_2^{(i)}) \cp \cp \nn \\
&+& \ap \ap \nu_p^{(i)} K_0^{(i)}( 2F(2)-1)-2F(1) K_2^{(i)} \cq  +... \cg   \,.
\label{and}
\eeqn
Let us introduce a physical scale by means of the following redefinition of the
$\mu^2$ parameter
\beq
 \log \frac{\mu^2}{|\La_p|}+F(2) \mapsto -\log(|\La|\mu^{-2})
\label{alfa}
\eeq

Hence, the leading part of $ \Ga^{(1)}$ is given by
\beqn
\Ga^{(1)}&=&\frac{\log (|\La|\mu^{-2})}{4}
\sum_{p,i}C_p^{(i)}
\ag  a^4 |\La_p^{(i)}|^2  K_0^{(i)} \cp \nn \\
&+& \ap
 a^{2}|\La_p^{(i)}|  (\nu_p^{(i)} K_0^{(i)}-K_2^{(i)})+...
\cg\, .
\label{byt}
\eeq

Now it is quite easy to rewrite the above one-loop corrections in
terms of geometric quantities, appearing in the classical action.
To this aim, it is sufficient to observe that, for constant curvature
space, we have
\beqn
  a^4 &=&\int d^4x \sqrt g \at Vol(\ca F_N) \ct^{-1}\,, \nn \\
  a^{2}&=&\int d^4x \sqrt g \kappa R\at 12 Vol(\ca F_N)
\ct^{-1}\,,
\label{corr}
\eeqn
where $\kappa=ka^2$, $\ca F_N $ is the fundamental domain of the
compact hyperbolic manifold

As a result,  using the relevant part of  $\Ga^{(1)} $ and the classical
action, one can write the one-loop effective action in the
large-distance limit as
\beqn
\Ga_{eff}&=& S+\Ga^{(1)}= \int d^4x \sqrt g \aq
 \La (8\pi G)^{-1} +   \be_{\La}|\La|^{2}\log  (|\La|\mu^{-2}) \cq \nn \\
&-& \int d^4x \sqrt g R \aq
  (16\pi G)^{-1} +   \be_{G}|\La|\log (|\La|\mu^{-2}) \cq \,,
\label{andrei}
\eeqn
where
\beq
\be_{\La}=\frac{1}{4 V(\ca F_N)}
\sum_{p,i} (b_p^{(i)})^{2}C_p^{(i)}  K_0^{(i)}\,,
\label{G}
\eeq

\beq
\be_{G}=-\frac{\kappa}{24 V(\ca F_N)}
\sum_{p,i} b_p^{(i)} C_p^{(i)} ( \nu_p^{(i)} K_0^{(i)}-K_2^{(i)})\,.
\label{G1}
\eeq
The effective Newton and cosmological constants turn out to be
\beq
\La_{eff} =\La \frac{1+\kappa \be_{\La} 8\pi G|\La| \log (
|\La|\mu^{-2})}{1+\be_{G} 16\pi G |\La| \log (|\La|\mu^{-2})}\,,
\label{laeff1}
\eeq

\beq
(G\La)_{eff} =(G \La )\frac{1+\kappa \be_{\La} 8\pi G|\La| \log (
|\La|\mu^{-2})}{\aq 1+\be_{G} 16\pi G |\La| \log (|\La|\mu^{-2})
\cq^2}\,.
\label{laeff2}
\eeq

In order to apply the general formula, we have to compute the
Seeley-deWitt cofficients $K_{2r}^{(i)}$, associated with the (constrained)
 spin-fields on an  compact constant curvature background, and this
could be done making use of the general algorithm.  In the case of compact
hyperbolic background $H^4/\Ga$, we may proceed  as following.

For scalar (spinor) fields,
if we limit ourselves to a smooth manifolds (strictly hyperbolic
subgroup of $\Ga$), one can use the Selberg trace formula. In the case
of scalar fields we have (see \cite{byts94u-325} for details)
\beqn
\Tr e^{-tL^{(0)}}&=& V(\ca F_4)  \int_{-\ii}^{\ii}
 \frac{ r (r^2+\fr{1}{4})\tanh \pi r}{8\pi^2} \, e^{-(r^2+\fr{9}{4})t} dr \nn
\\
&+&\sum_{\ag \ga \cg}\sum_{n=1}^\ii \frac{\chi(\ga)^n l_\ga}{S_4
(n,l_\ga)}
\frac{1}{\sqrt{4\pi t}}\exp {\aq -(\fr{9}{4}t+\fr{n^2l_\ga^2}{4t})\cq}\,.
\label{5.60}
\eeqn
Above we have the first term (identity contribution) and the
topological one, wich however is exponentially small for $ t \to 0$.
Thus the heat kernel coefficients are contained in the identity term,
which are the contribition related to $H^4$, apart the volume
normalization. This means that the contribution which are interested in
are computable, for higher spins, from the exact heat-kernel for the
Laplace type operator on $H^4$. This heat-kernel can be computed using the
result on the constrained higher-spin Plancherel measure $\mu^{(s)}(r)
 $ on $H^4$,
evaluated in Ref.\cite{camp93-47-3339}. The heat-kernel trace can be
written as
\beq
K(t|L^{(s)})=V(\ca F_4) e^{-t(\fr{9}{4}+s)}\int_0^\ii
 e^{-t r^2}\,\mu^{(s)}(r)\,dr
\eeq
where
\beq
\mu^{(s)}(r)=\frac{(2s+1)}{8\pi^2}\aq (r^2+(s+\fr{1}{2})^2\cq \tanh
\pi(r+is)\,.
\label{P}
\eeq
Making use of
\beq
\tanh \pi(r+is)=\ag
\begin{array}{cc}
1-\frac{2}{1+e^{2\pi r}}&\,, \,\,\, s=0,1,2,...\\
1-\frac{2}{1-e^{2\pi r}}& \,, \,\,\, s=1/2,3/2,...
\end{array}
\cp
\label{}
\eeq
and the identities
\beqn
\int_0^\ii \frac{r^{2n-1}}{e^{2\pi r}+1} \,dr&=&
\frac{(-1)^{n-1} (1-2^{1-2n})}{4n} B_{2n} \nn \\
\int_0^\ii \frac{r^{2n-1}}{e^{2\pi r}-1}\, dr&=&
\frac{(-1)^{n-1} }{4n} B_{2n}
\eeqn
in which $B_{2n}$ are the Bernoulli numbers, one obtains the first
Seeley-DeWitt coefficients related to spin-constrained Laplacian
$L^{(s)}$:

For $s=0,1,2,..$
\beqn
K^{(s)}_0&=&\frac{(2s+1)}{(4\pi)^2}V(\ca F_4) \nn \\
K^{(s)}_2&=&\frac{(2s+1)}{(4\pi)^2}V(\ca F_4)(s^2-2)=(s^2-2)(2s+1)K^{(0)}_0
\nn \\
K^{(s)}_4&=&\frac{(2s+1)}{(4\pi)^2} V(\ca F_4)\aq
(\fr{9}{4}+s)(\fr{7}{8}-s^2-\fr{s}{2})-
\fr{1}{12}(s+\fr{1}{2})^2+\fr{7}{40}   \cq
\label{ke}
\eeqn

For $s=1/2,3/2,...$
\beqn
K^{(s)}_0&=&\frac{(2s+1)}{(4\pi)^2}V(\ca F_4) \nn \\
K^{(s)}_2&=&\frac{(2s+1)}{(4\pi)^2}V(\ca
F_4)(s^2-2)=(s^2-2)(2s+1)K^{(0)}_0 \nn \\
K^{(s)}_4&=&\frac{(2s+1)}{(4\pi)^2}V(\ca F_4)\aq
(\fr{9}{4}+s)^2-(\fr{9}{2}+2s-\fr{1}{3})(s+\fr{1}{2})^2+\fr{1}{30}   \cq
\label{ko}
\eeqn

{}From the above equations,
we obtain
\beq
\be_{\La}=\frac{1}{4V(\ca F_4)}\sum_{p,i} (b_p^{(i)})^2 C_p^{(i)}  K_0^{(i)}=
= \frac{1}{4(4\pi)^{2}}\sum_{p,i} (2i+1)(b_p^{(i)})^2 C_p^{(i)} \,,
\label{b1}
\eeq
\beq
 \be_G=\frac{1}{24V(\ca F_4)}\sum_{p,i} b_p^{(i)}C_p^{(i)}
 (\nu_pK_0^{(i)}- K_2^{(i)})=
  \frac{1}{(24\pi)^{2}} \sum_{p,i} b_p^{(i)}C_p^{(i)}(2i+1)
 (\nu_p^{(i)}-i^2+2)\, ,
\label{g1}
\eeq

The main interest in using a supergravity theory stems from the fact that the
bosonic and fermionic degrees of freedom enter the one-loop path integral with
opposite contributions ($C_p^{(s)}$ have opposite signs). Thus, there
might be the possibility to have a cancellation in the $\be_\La$
expression, which is impossible in a pure gravity theory
\cite{byts94r}. This is also supported by the improved one-loop ultraviolet
behaviour related to every supergravity theory.

Our next aim will be to illustrate these general considerations to the
 O(4) gauged supergravity.
The Lagrangian  for this theory is given by
\cite{das,frad84-234-472}
\beqn
L&=&-\frac{R}{K}+\frac{2}{K^2}\frac{\partial_\mu \Phi \partial^\mu
\Phi^*}{\at 1-|\Phi|^2 \ct ^2}-\frac{8 g^2}{K^4}\at 1+
\frac{2}{1-|\Phi|^2}\ct \nn \\
&+&\frac{1}{8}\at F_{\mu \nu}^{ij} \ct^2+\ag \frac{1}{8}\frac{\Phi}{1-|\Phi|^2}
\at \Phi \de_{ik}\de_{jl}-\frac{1}{2}\epsilon_{ijkl}\ct F_{\mu \nu}^{ij}
\at F_{\mu \nu}^{kl} +i\tilde{F}_{\mu \nu}^{kl}  \ct+h.c. \cg \nn \\
&+&\frac{1}{2}\epsilon^{\mu \nu \rho \sigma}\bar{\Psi}^i_\mu
\ga_5\ga_\nu D_\rho \Psi_\sigma^i+\frac{2g}{K\sqrt {\at
1-|\Phi|^2\ct}}\bar{\Psi}^i_\mu\sigma_{\mu \nu}\Psi^i_\nu \nn \\
&+&\frac{1}{2}\bar{\chi}^i\ga_\mu D_\mu\chi^i+\frac{\sqrt 2 g}{K\sqrt {\at
1-|\Phi|^2\ct}}\bar{\Psi}^i_\mu \ga_\mu\at
\Phi_1+i\ga_5\Phi_2\ct\chi^i\nn \\
&+&( \mbox{quartic fermionic terms})\,,
\label{sg}
\eeqn
where $K^2=16\pi G$, $\sigma_{\mu \nu}=\fr{1}{2}[\ga_\mu,\ga_\nu]$,
$\ag \ga_\mu,\ga_\nu \cg=g_{\mu \nu}$,  $\ga_5^2=1$ , $\tilde{ F_{\mu
\nu}}=\fr{1}{2}\epsilon^{\mu \nu \rho \sigma}F_{\rho \sigma}$ and
$D_\mu$ is the total covariant gravitational and gauge derivative. The
physical content of the fields are: the gravitons, $A_\mu^i$ $O(4)$ gauge
fields $(i,j=1,...4)$  ($ F_{\mu \nu}^{kl}$ being the field strenght), a
complex scalar field $\Phi= \Phi_1+i\Phi_2$, satisfying
the constraint $|\Phi|<1$,   four $\Psi_\mu^i$ (Majorana gravitinos) and
four $\chi^i$ (Majorana spinors). The related action is invariant under gauged
$N=4$
supersymmetry (for details see \cite{das,brei,frad84-234-472}).
We will make the one loop-approximation expanding all the fields
around the stable supersymmetric vacuum \cite{frad84-234-472}
\beq
\Phi=0\,,\Psi=0\,, A=0\,, \chi=0\,, \La_{\mbox{on
shell}}=\La_0=-\frac{12g^2}{K^2}
\label{V}
\eeq
thus $g_{\mu \nu}$ correspond to anti-de Sitter space and the
tree-level
cosmological constant turns out to be negative, namely
\beq
\La=-\frac{g^2}{K^2}\at 1+2\al \ct\, \hs \al=\at 1-|\Phi|^2 \ct^{-1}
\label{a}
\eeq
Note that $\Phi=0$ is the only extremum (maximum) of the classical
potential (third term in Eq. (\ref{sg})).

The one-loop Euclidean effective action on the background (\ref{V}) has been
evaluated (making use of the De Donder gauge for the graviton field
and $\Psi=0 $ gauge  for the gravitino) in Ref. \cite{frad84-234-472}. The
result reads
\beqn
\Ga^{(1)}&=&\frac{1}{2} \log \det \at -\nabla_{2}^2+2|\La|-\frac{8}{a^2} \ct
-\frac{1}{2} \log \det \at -\nabla_{1}^2+\frac{3}{a^2} \ct+
3\log \det \at -\nabla_{1}^2-\frac{3}{a^2} \ct   \nn \\
&+& \frac{1}{2} \log \det \at -\nabla_{0}^2-4|\La|+\frac{12}{a^2} \ct
- \frac{7}{2}\log \det \at -\nabla_{0}^2\ct
+ \frac{1}{2} \log \det \at -\nabla_{0}^2-2|\La|\frac{4\al-3}{1+2\al}
\ct\nn\\
&+&  \frac{1}{2} \log \det \at -\nabla_{0}^2-2|\La|\frac{1}{1+2\al} \ct
-  \log \det \at -\nabla_{3/2}^2+|\La|\frac{\al}{1+2\al} \ct \nn \\
&+&\log \det \at -\nabla_{1/2}^2+4|\La|\frac{\al}{1+2\al} \ct
-\log \det \at -\nabla_{1/2}^2 \ct\,,
\label{dau}
\eeqn
in which the $0,1/2, 1, 3/2, 2$ are labelling the scalar, spinor, tranverse
vector, transverse gravitino and
traceless transverse tensor respectively.

In our approach, all the $O^{(i)}_p$ operators should be positive
definite. One can show that in the scalar sector, some operators are
negative definite. Furthermore, the indefiniteness may  depend on
the  gauge fixing. There exist several possibilities to handle this
problem. One of these might consist in considering a suitable
analytical continuation, when necessary, from $-|\La|$ to $|\La|$.
This regularization procedure gives finite results,  but the price to
pay is the appearance of imaginary terms in the effective action,
signalling a quantum metastability. However, in the large-distance
limit, we may formally consider some $b_p^{(s)}<0$, without any
problem, the imaginary terms being non leading.

Keeping in mind such observations, Eq. (\ref{dau}) gives

\beq
\be_\La=\frac{1}{(4\pi)^2}\frac{66\al^2+12\al+19}{(1+2\al)^2}
\eeq
and

\beq
\be_G=-\frac{1}{24(4\pi)^2}\frac{185\al+76}{(1+2\al)}\,,
\eeq
which are an example of computation.

Finally we  would like briefly to comment on the issue related to
richer geometric structure we are dealing with. In the explicit 4-dimensional
example presented for illustrative purposes, we have considered a
smooth compact hyperbolic manifold $H^4/\Ga$. Technically this is
equivalent to the assumption that only hyperbolic elements are
present. However, it is known that also elliptic elements may be
present, and their contribution can be taken into account by means of
the Selberg trace formula. It is expected an elliptic correction to
$\be_G$ of the form \cite{byts94r}

\beq
\be_G=-\frac{1}{24(4\pi)^2}\aq \frac{185\al+76}{(1+2\al)}+
\sum_{p,i} b_p^{(i)}C_p^{(i)}\frac{K_{2,E}^{(i)}}{K_0^{(0)}} \cq\,,
\eeq
where $K_{2,E}^{(i)}$ are the contributions related to the elliptic
elements of $\Ga$.

In conclusion, we have discussed the one-loop effective action for
$O(4)$ supergravity with negative cosmological constant on a
hyperbolic background, by making use of zeta-function regularization
and heat-kernel techniques.
The use of large-distant limit approximation has permitted to obtain
reasonable simple expressions for the effective one-loop cosmological
and gravitational constants. These expressions, in the large-distance
limit, depend on the heat-kernel coefficients, which can be computed.
A novel feature, with
respect to the spherical ($\La>0$) background, consists in the richer
geometric structure one has to deal with. As a consequence, the value
of the coefficient $\be_G$ may also depends on  the choice of the
topological non-trivial field configurations (twisted or untwisted
fields, see also \cite{avis} for discussion of twisted fields
on toroidal spaces) on the orbifolds $H^4/\Ga$.

\ack{ We thank G. Cognola and L. Vanzo for discussions. A.A. Bytsenko wishes
to thank INFN and the Department of Physics of
Trento University for financial support and kind hospitality. S.D.
Odintsov thanks the Generalitat de Catalonya for financial support. }


\begin{thebibliography}{10}


\bibitem{gibb78-138-141}
G.W.~Gibbons, S.W.~Hawking and M.J.~Perry.
Nucl.~Phys., {\bf B138}, 141, (1978).

\bibitem{gibp}
 G.W.~Gibbons and  M.J.~Perry.
Nucl.~Phys.,{\bf B146}, 90, (1978).

\bibitem{hawk79b}
S.W.~Hawking.
The path-integral approach to quantum gravity.
In {\it General Relativity.~An Einstein Centenary Survey}, S.W.~Hawking and
  W.~Israel, editors.
Cambridge University Press, Cambridge, (1979).

\bibitem{chri80-170-480}
S.M.~Christensen and M.J.~Duff.
Nucl.~Phys., {\bf B170}, 480, (1980).

\bibitem{cdrg}
S.M.~Christensen, M.J.~Duff, G.W.~Gibbons and M.~Rocek.
Phys.~Rev.~Lett., {\bf 45}, 161, (1980).

\bibitem{anti}
I.~Antoniadis, J.~Iliopulos and T.N.~Tomaras.
Phys.~Rev.~Lett., {\bf 56}, 1319, (1986).

\bibitem{ford}
L.~Ford.
Phys.~Rev., {\bf D31}, 710 (1985).

\bibitem{allen}
B.~Allen and M.~Turyn.
Nucl.~Phys., {\bf B292}, 813, (1987).


\bibitem{odin89}
S.~D.~Odintsov.
Europhys.~Lett., {\bf 10}, 287, (1989);
Theor. Math. Phys., {\bf 82}, 61, (1990).

\bibitem{antm}
I.~Antoniadis and E.~Mottola.
J.~Math.~Phys., {\bf 32}, 1037, (1991).

\bibitem{frad84-234-472}
E.S.~Fradkin and A.A.~Tseytlin.
Nucl.~Phys., {\bf B234}, 472, (1984).

\bibitem{tayl90-345-210}
T.R.~Taylor and G.~Veneziano.
Nucl.~Phys., {\bf B345}, 210, (1990).

\bibitem{buch}
I.L.~Buchbinder, S.D.~Odintsov and I.L.~Shapiro.
{\it Effective action in quantum gravity}.
IOP Publishing, Bristol and Philadelphia, (1992).


\bibitem{byts94r}
 A.A.~Bytsenko, S.D.~Odintsov and and S.~Zerbini.
Trento University preprint UTF 328, (1994).

\bibitem{hawk77-55-133}
S.W.~Hawking.
Commun.~Math.~Phys., {\bf 55}, 133, (1977).

\bibitem{dowk76-13-3224}
J.~Dowker and R.Critchley.
  Phys.~Rev. {\bf D13}, 3224,(1976).

\bibitem{eliz}
E.~Elizalde, S.D.~Odintsov, A.~Romeo, A.A.~Bytsenko and S.~Zerbini.
Zeta-regularization with applications.World Sci., Singapore, 1994.

\bibitem{byts94u-325}
A.A.~Bytsenko, G.~Cognola, L.~Vanzo and S.~Zerbini.
Quantum fields and extended objects in space-times with constant curvature
  spatial section.
 Trento University Preprint UTF 325, (1994).


\bibitem{polc}
J. Polchinski
Phys.~Lett., {\bf B219}, 251, (1989).


\bibitem{camp93-47-3339}
R. Camporesi and A.~Higuchi.
Phys. Rev., {\bf D47}, 3339, (1993).


\bibitem{das}
A.~Das,M. Fisler and M.~Rocek.
{}~Phys. Rev., {\bf D16}, 3427, (1977);
B. de Wit and H. Nicolai.
Nucl. Phys., {\bf B188},98, (1981);Nucl. Phys., {\bf B208},323, (1982)

\bibitem{brei}
P. Breitenlohner  and D. Z. Freedman.
Ann..~Phys., {\bf 144}, 249, (1982);
G.W. Gibbons, C. M. Hull and N.P. Warner
Nucl.~Phys., {\bf B218}, 173, (1983).

\bibitem{avis}
S.J.Avis and C.J.Isham.
Nucl.~Phys.,{\bf B156}, 441, (1979).


\end{thebibliography}
\end{document}